\documentclass[10pt, a4paper,conference]{IEEEtran}  
\IEEEoverridecommandlockouts
\usepackage{cite}
\usepackage{amsmath,amssymb,amsfonts}
\usepackage{algorithmic}
\usepackage{graphicx}
\usepackage{tabularx}
\usepackage{siunitx}
\usepackage{listings}
\usepackage{textcomp}
\usepackage{xcolor}
\def\BibTeX{{\rm B\kern-.05em{\sc i\kern-.025em b}\kern-.08em
    T\kern-.1667em\lower.7ex\hbox{E}\kern-.125emX}}

\usepackage{todonotes}
\usepackage[utf8]{inputenc}
\usepackage{listingsutf8}
\usepackage{tabularx}
\usepackage{tabularray}
\usepackage[inkscapearea=page]{svg}
\usepackage{adjustbox}
\usepackage{tabularx}
\usepackage[para,online,flushleft]{threeparttable}
\usepackage{multirow}
\usepackage{amssymb}
\usepackage{pifont}
\usepackage{booktabs}
\usepackage{float}
\usepackage[justification=centering]{caption}

\usepackage{hyperref}

\ttfamily

\definecolor{mGreen}{rgb}{0,0.6,0}
\definecolor{mGray}{rgb}{0.5,0.5,0.5}
\definecolor{mPurple}{rgb}{0.58,0,0.82}
\definecolor{backgroundColour}{rgb}{0.95,0.95,0.92}

\lstdefinestyle{CStyle}{
    backgroundcolor=\color{white},   
    commentstyle=\color{mGreen},
    keywordstyle=\bfseries,
    numberstyle=\tiny\color{mGray},
    stringstyle=\itshape,
    morekeywords={omp},
    otherkeywords={\#pragma},
    keywords=[2]{target, device, map, to, asm},
    keywordstyle=[2]{\underbar},
    frame=bt,
    numbers=left,
    xleftmargin=2em,
    framexleftmargin=1.5em
}

\lstdefinestyle{my_dts}{
  literate=
    {├}{{\smash{\raisebox{-1ex}{\rule{1pt}{\baselineskip}}}\raisebox{0.5ex}{\rule{1ex}{1pt}}}}1 
    {─}{{\raisebox{0.5ex}{\rule{1.5ex}{1pt}}}}1 
    {└}{{\smash{\raisebox{0.5ex}{\rule{1pt}{\dimexpr\baselineskip-1.5ex}}}\raisebox{0.5ex}{\rule{1ex}{1pt}}}}1
    {│}{{\smash{\raisebox{-1ex}{\rule{1pt}{\baselineskip}}}\raisebox{0.5ex}{\rule{1ex}{0pt}}}}1,
    emph={ranges},
    emphstyle={\textit},
    numbers=none,
}

\usepackage[toc, section=section]{glossaries}
\makeglossaries

\newacronym{abi}{ABI}{application binary interface}
\newacronym{ai}{AI}{artificial intelligence}
\newacronym{api}{API}{application programming interface}
\newacronym{apu}{APU}{accelerated processing unit}
\newacronym{asic}{ASIC}{application-specific integrated circuit}
\newacronym{axi}{AXI}{advanced extensible interface}
\newacronym{dsa}{DSA}{domain-specific accelerator}

\newacronym{ptw}{PTW}{page table walk}
\newacronym[plural=PTEs, longplural=page table entries]{pte}{PTE}{page table entry}

\newacronym{rtt}{RTT}{round-trip time}

\newacronym{cccc}{CCCC}{C and C++ Code Counter}
\newacronym{cnn}{CNN}{convolutional neural network}
\newacronym{cpu}{CPU}{central processing unit}
\newacronym{c2c}{C2C}{chip-to-chip}

\newacronym{dma}{DMA}{direct memory access}
\newacronym{dram}{DRAM}{dynamic random access memory}

\newacronym{dnn}{DNN}{deep neural network}

\newacronym{eoc}{EOC}{end of computation}
\newacronym{epac}{EPAC}{European processor accelerator}

\newacronym{fdt}{FDT}{flattened device tree}
\newacronym{fpga}{FPGA}{field-programmable gate array}
\newacronym{flops}{FLOPs}{floating-point operations per second}
\newacronym{fpu}{FPU}{floating-point unit}

\newacronym{gpu}{GPU}{graphical processing unit}

\newacronym{hart}{hart}{hardware thread}
\newacronym[plural=heSoCs, longplural=heterogeneous systems-on-hip]{hesoc}{heSoC}{heterogeneous system-on-chip}
\newacronym{hpc}{HPC}{high performance computing}
\newacronym{hls}{HLS}{high level synthesis}

\newacronym{idol}{I\$}{instruction cache}
\newacronym{iommu}{IOMMU}{IO memory management Unit}
\newacronym{iotlb}{IOTLB}{IO translation lookaside buffer}
\newacronym{iova}{IOVA}{IO virtual address}
\newacronym{isa}{ISA}{instruction set architecture}

\newacronym{lsu}{LSU}{load store unit}
\newacronym{llc}{LLC}{last-level cache}

\newacronym{mac}{MAC}{multiply and accumulate}
\newacronym{mimd}{MIMD}{multiple-instructions-multiple-data}
\newacronym{ml}{ML}{Machine Learning}
\newacronym{mmu}{MMU}{memory management unit}
\newacronym{mpsoc}{MPSoC}{multi-processor SoC}

\newacronym{loc}{LOC}{lines of code}
\newacronym[longplural={networks-on-chip}]{noc}{NoC}{network-on-chip}

\newacronym{os}{OS}{operating system}

\newacronym{rab}{RAB}{remapping address block}

\newacronym{pci}{PCI}{peripheral component interconnect}
\newacronym{pcie}{PCIe}{peripheral component interconnect express}
\newacronym{pe}{PE}{processing element}
\newacronym{pmca}{PMCA}{programmable manycore accelerator}

\newacronym{sdk}{SDK}{software development kit}
\newacronym{simd}{SIMD}{single-instruction-multiple-data}
\newacronym{simt}{SIMT}{single-instruction-multiple-threads}
\newacronym{slc}{SLC}{system-level cache}
\newacronym[plural=SoCs, longplural=systems-on-chip]{soc}{SoC}{system-on-chip}
\newacronym[longplural={scratchpad memories}]{spm}{SPM}{scratchpad memory}
\newacronym{spmd}{SPMD}{single program multiple data}
\newacronym{sram}{SRAM}{static random access memory}

\newacronym{tcdm}{TCDM}{tightly-coupled data memory}
\newacronym{tlb}{TLB}{translation lookaside buffer}

\begin{document}


\def\reviewpass{v0.0.1}
\def\showrevision{}

\ifx\showrevision\undefined
    \newcommand{\todoo}[1]{{#1}}
    \newcommand{\ph}[1]{{#1}}
\else
    \newcommand{\todoo}[1]{{\textcolor{red}{#1}}\PackageWarning{TODO:}{#1!}}
\fi

\title{Evaluating IOMMU-Based Shared Virtual Addressing for RISC\nobreakdash-V Embedded Heterogeneous SoCs 
}

\ifx\blind\undefined
\author{\IEEEauthorblockN{Cyril Koenig}
\IEEEauthorblockA{\textit{Integrated Systems Laboratory} \\
\textit{ETH Zurich}\\
cykoenig@iis.ee.ethz.ch}
\and
\IEEEauthorblockN{Enrico Zelioli}
\IEEEauthorblockA{\textit{Integrated Systems Laboratory} \\
\textit{ETH Zurich}\\
ezelioli@iis.ee.ethz.ch}
\and
\IEEEauthorblockN{Luca Benini}
\IEEEauthorblockA{
\textit{ETH Zurich} \\
Zurich, Switzerland \\
\textit{Università di Bologna} \\
Bologna, Italy \\
lbenini@iis.ee.ethz.ch}
}
\else
    \author{%
            \vspace{0.4cm} %
            \textit{Authors omitted for blind review}
            }
\fi

\maketitle

\begin{abstract}

Embedded heterogeneous \glspl{soc} rely on domain-specific hardware accelerators to improve performance and energy efficiency. In particular, programmable multi-core accelerators feature a cluster of processing elements and tightly coupled scratchpad memories to balance performance, energy efficiency, and flexibility. In embedded systems running a general-purpose OS, accelerators access data via dedicated, physically addressed memory regions. This negatively impacts memory utilization and performance by requiring a copy from the virtual host address to the physical accelerator address space. Input-Output Memory Management Units (IOMMUs) overcome this limitation by allowing devices and hosts to use a shared virtual paged address space. However, resolving IO virtual addresses can be particularly costly on high-latency memory systems as it requires up to three sequential memory accesses on IOTLB miss.

In this work, we present a quantitative evaluation of shared virtual addressing in RISC\nobreakdash-V heterogeneous embedded systems. We integrate an IOMMU in an open source heterogeneous RISC\nobreakdash-V SoC consisting of a 64\nobreakdash-bit host with a 32\nobreakdash-bit accelerator cluster. We evaluated the system performance by emulating the design on FPGA and implementing compute kernels from the RajaPERF benchmark suite using heterogeneous OpenMP programming. We measure the transfers and computation time on the host and accelerators for systems with different DRAM access latencies. We first show that IO virtual address translation can account for $4.2\%$ up to $17.6\%$ of the accelerator’s runtime for \texttt{gemm} (General Matrix Multiplication) at low and high memory bandwidth. Then, we show that in systems containing a last-level cache, this IO address translation cost falls to $0.4\%$ and $0.7\%$ under the same conditions, making shared-virtual addressing and zero-copy offloading suitable for such RISC\nobreakdash-V heterogeneous SoCs.

\end{abstract}

\begin{IEEEkeywords}
RISC\nobreakdash-V, Hardware accelerators, Heterogeneous computing, IOMMU, Shared virtual addressing
\end{IEEEkeywords}

\section{Introduction}

Heterogeneous \glspl{soc} employ domain-specific accelerators to improve performance and energy efficiency. Vendors such as Nvidia and AMD have developed platforms coupling integrated \gls{gpu} in so-called \gls{apu}, offering high performance from the edge - with the Nvidia Tegra series~\cite{nvidia_xavier_2019} - to datacenters - with the AMD EPYC processors~\cite{amd_exa_2023}. In these platforms, the host and accelerator share the same memory controller, making them particularly suited to also share a unified virtual address space to simplify data sharing in heterogeneous applications. Traditionally, such systems enable shared virtual memory using a hardware IO-\gls{mmu} located between the highest \gls{gpu} cache and the memory fabric~\cite{nvidia_xavier_2019}.

With the democratization of the open source RISC\nobreakdash-V \gls{isa}, a new era of heterogeneous platforms opens, combining RISC\nobreakdash-V hosts and RISC\nobreakdash-V accelerator IPs implementing custom extensions. Similarly to \glspl{apu}, these new platforms can also leverage shared virtual addressing and zero-copy offloading using the recently ratified RISC\nobreakdash-V \gls{iommu} specification 1.0~\cite{riscv_iommu}.

To this day, many of the proposed RISC\nobreakdash-V parallel accelerator architectures~\cite{esperanto_soc_2022}~\cite{tenstorrent_2024}~\cite{occamy_2024} follow the \gls{pmca} design pattern. These chips consist of multiple \glspl{pe} assembled in clusters around a fast-access \gls{spm} refilled using explicit \gls{dma} calls.

In this work, we integrate the open source RISC\nobreakdash-V  \gls{iommu} implementation from~\cite{pinto_iommu_2023}~\cite{zerodaylabs_github} into a heterogeneous open source \gls{soc} combining a RISC\nobreakdash-V Linux capable host and a floating-point optimized RISC\nobreakdash-V \gls{pmca}. By emulating this platform on FPGA, we evaluate the accelerator's performance on four heterogeneous compute benchmarks using heterogeneous OpenMP-offloading. We show a significant increase in data transfer time when enabling shared virtual memory support, causing up to $17.6\%$ performance degradation on \texttt{gemm} at high memory latency. However, we show that after integrating a shared \gls{llc} before the memory controller, degradations caused by \gls{iotlb} misses get down to $0.7\%$ without requiring any further \gls{iommu} optimization or extensions~\cite{multi_lvl_tlb_2017}~\cite{iotlb_coalescing_2018}. We then confirm these results under synthetic concurrent memory traffic on the shared cache from the host.

Overall, we present the following contributions:

\begin{itemize}
  \item The integration of the \gls{iommu} from~\cite{pinto_iommu_2023}~\cite{zerodaylabs_github} in a RISC\nobreakdash-V heterogeneous \gls{soc} featuring an eight-cores scratchpad-based \gls{pmca}. The hardware is available in open source\footnote[1]{\href{https://github.com/pulp-platform/carfield/tree/date_iommu_evaluation}{https://github.com/pulp-platform/carfield/tree/date\_iommu\_evaluation}}.
  \item The implementation and benchmarking of heterogeneous userspace applications on an FPGA demonstrator under different memory latencies.
  \item The evaluation of shared virtual addressing benefits on this demonstrator without any custom \gls{iommu} architectural optimization when coupled to an \gls{llc}.
\end{itemize}

\glsunset{tlb}
\glsunset{mmu}


\section{Background}
This study builds upon open-source architectural building blocks and integrates several hardware and software components to propose an evaluation of shared virtual addressing on heterogeneous RISC\nobreakdash-V platforms.

Cheshire~\cite{cheshire_2023} is a fully open source and an in-depth host \gls{soc} built around the 64-bit, application-class CVA6 RISC\nobreakdash-V processor~\cite{cva6_2019}. Alongside the CVA6 core, Cheshire offers multiple peripherals for IO communication and an \gls{llc} that can be partitioned between cache and \gls{sram} via run-time configurations. The system components are connected via an \gls{axi} fully connected crossbar~\cite{pulp_axi4_2022}. The Cheshire \gls{soc} supports the seamless integration of \glspl{dsa}, which can be connected to the system's AXI crossbar via dedicated master and slave ports. 

The Snitch cluster\cite{snitch_2020} is an open source, RISC\nobreakdash-V, multi-core accelerator targeting energy-efficient execution of floating-point workloads. The cluster comprises eight \texttt{rv32imafd} \glspl{pe}, which are small Snitch integer cores coupled to FPUs. A ninth core controls the cluster's \gls{dma} engine. The cluster contains an L1 \gls{tcdm} split into parallel accessible banks. In this work, the Snitch cluster is referred to as \emph{device} or \emph{accelerator} interchangeably.

The RISC\nobreakdash-V \gls{iommu} specification v1.0~\cite{riscv_iommu} defines the hardware and software standards to build \gls{iova} support for devices and \gls{dma} engines. This specification was recently implemented in an open source IP~\cite{pinto_iommu_2023} available at~\cite{zerodaylabs_github}. This module can enable memory protection, virtualization, and zero-copy heterogeneous applications.

OpenMP~\cite{openmp_spec_v5} is a standardized \gls{api} for parallel C/C++ and Fortran programming. It also supports heterogeneous programming via its \emph{target} \gls{api}. In this work, we use this \gls{api} to build userspace applications accelerated by a Snitch cluster \gls{dsa}.


\section{Methods}\label{sec:platform}

\begin{figure}[t]
    \centering
    \includegraphics[width=0.9\columnwidth]{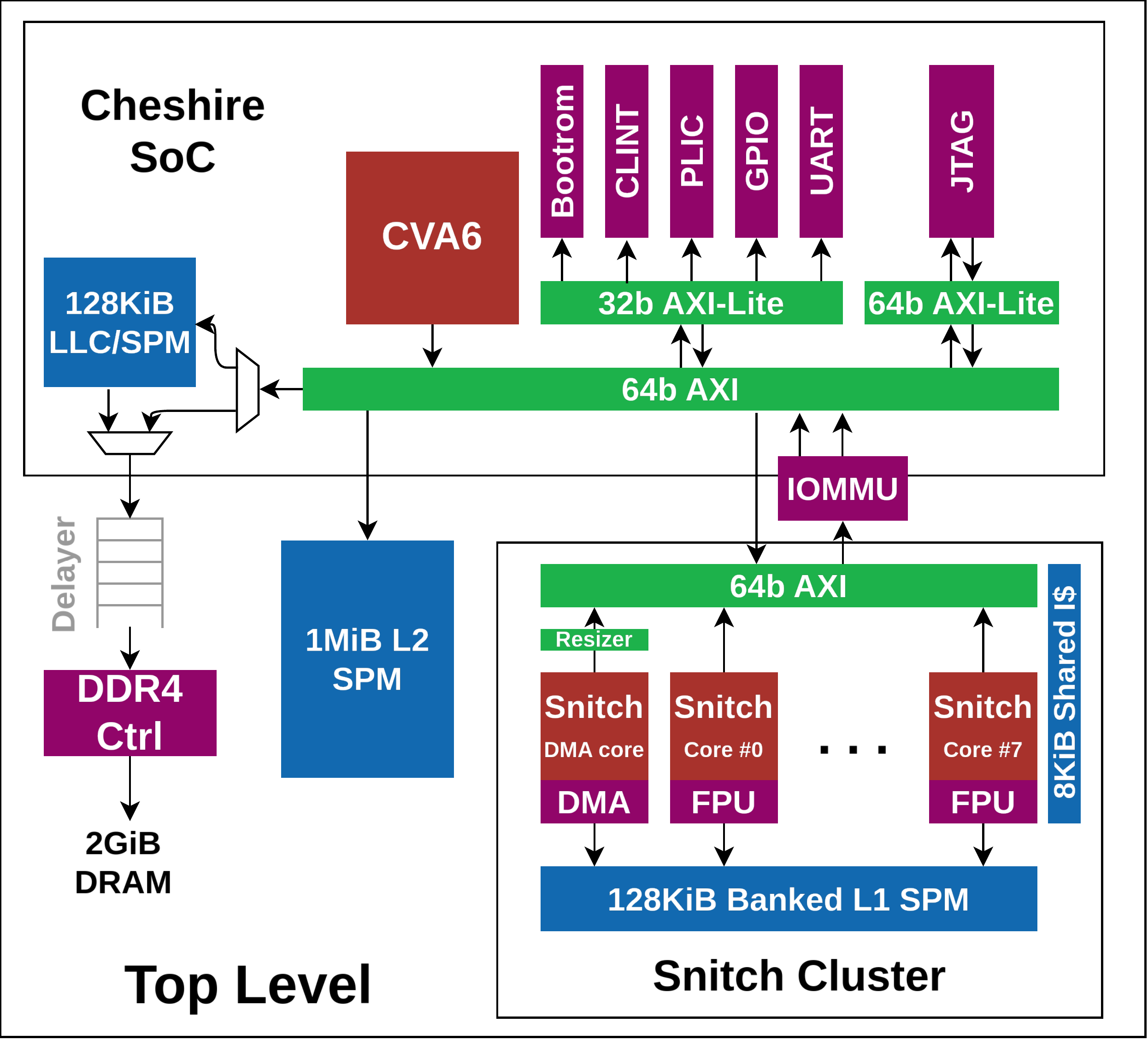}
    \caption{Block diagram of the prototype platform. Note that the DRAM delayer is added on emulation to provide more accurate performance evaluation.}
    \label{fig:archi:soc}
\end{figure}

\subsection{Hardware and software architecture}\label{subsec:archi:block_diagram}

Based on the key building blocks described above, we designed a prototype heterogeneous platform to explore shared virtual addressing on RISC\nobreakdash-V. The proposed platform is presented in Figure~\ref{fig:archi:soc}, the Cheshire \gls{soc} is configured  with \SI{128}{\kibi\byte} of \gls{llc}/\gls{spm} and a single-core 64\nobreakdash-bit CVA6 with \SI{32}{\kibi\byte} of write-through data-cache. The \gls{iommu} is located between the \gls{soc} crossbar interconnect and the Snitch cluster. It is configured with four \gls{iotlb} entries and one device directory entry, caching the location of the root page table for one (device, process) pair. The \gls{iommu} connects to the main crossbar via two \gls{axi} ports, one for the device transactions and one for \gls{ptw} transactions. The platform also contains \SI{1}{\mebi\byte} of non-cached and physically addressed scratchpad memory (one-to-one translation). This L2 \gls{spm} stores device binaries and shared data structures such as software mailboxes for synchronization. Finally, the DDR controller is connected to \SI{2}{\gibi\byte} of off-chip \gls{dram}, the lower half is allocated to Linux, the upper half is reserved for allocating physically contiguous \gls{dma} buffers for the accelerator in case the application does not use shared virtual memory but explicit data copies. The reserved \gls{dram} space is uncached by the \gls{llc} via muxes visible on the left of the system crossbar on Figure~\ref{fig:archi:soc}.

The platform is emulated on a Xilinx Ultrascale+ VCU128 \gls{fpga} development board, the Snitch cluster domain is clocked at \SI{20}{\mega\hertz} while the rest of the platform, including the \gls{iommu}, belongs to the host domain clocked at \SI{50}{\mega\hertz}. Note that multiple clock domains are common in heterogeneous \glspl{soc}; for instance, in the Nvidia Xavier chips, the \gls{cpu} frequency can be up to  $1.7\times$ the \gls{gpu} frequency~\cite{nvidia_xavier_2019}.
However, due to the low operating frequency of the emulated platform compared to the \gls{dram} chip, the main memory latency appears consequently smaller on \gls{fpga} than on silicon. At \SI{50}{\mega\hertz}, CVA6 has a read latency in \gls{dram} of only about 35 clock cycles. We add a parametrizable \gls{axi} delayer before the DDR controller to address this limitation. This module is implemented with FIFO macroblocks and delays \texttt{b} and \texttt{r} channels for a configurable number of cycles. This allows us to quantitatively evaluate the effects of main memory latency on the system's performance, especially when IOMMU-based address translation is enabled.

\subsection{Software stack and benchmarks}
The CVA6 core runs a patched Linux~6.9.0 with support for the RISC\nobreakdash-V \gls{iommu} Architecture Specification Version 1.0~\cite{linux_iommu_patch}. We implement a simple device driver to attach our accelerator to \gls{iommu} domain and create IO-virtual-physical mappings from the host. This driver also enables CVA6 to access the accelerator's register space, L2, and a reserved portion of the \gls{dram}. The driver can be accessed via a user-space library \gls{api}, which is then used by the heterogeneous OpenMP runtime to trigger kernel computation on the device.

In order to benchmark the performance of the accelerated system under different memory requirements, we implement a subset of linear and non-linear benchmarks from the RajaPERF suite~\cite{rajaperf_2019} presented in Table~\ref{tab:kern_desc}. For linear kernels, we select operations from the \emph{basic} and \emph{polybench} group with increasing arithmetic intensity: \texttt{axpy}, \texttt{heat3d}, \texttt{gesummv}, and \texttt{gemm}. For nonlinear kernels, we implement \texttt{sort} from the \emph{algorithm} group (using a parallel merge sort). All our benchmarks are run for single-precision floating-point data using input tiling and double-buffering to make the best use of the L1\nobreakdash-\gls{spm} and the cluster's \gls{dma} engine. In order to measure the \gls{iommu} effects for input and output data only, we offload each computation twice and measure performance with a hot instruction cache. In \glspl{soc} based on this evaluation platform, the cluster's instruction cache can bypass page translation with a different device identifier pointing to a bypassed device directory entry~\cite{riscv_iommu}. For comparison, we implement the same compute kernels on the single-threaded host core.

    

    
        
                
    
    
    
        
    
    
    
    

\begin{table}[t!]
    \caption{Total runtime in cycles for each kernel at variable memory latency.}
    \begin{center}
    \begin{tabular}{lcl}
    
        \toprule
        
        \textbf{Kernel} & \textbf{Input size} & \textbf{Description} \\  
                
        \midrule
        \texttt{gemm} & $128\times128$ &  Generic matrix-matrix multiplication. \\

        \texttt{gesummv} & $512\times512$ & Generic matrix-vector multiplication. \\
    
        \texttt{heat3d} & $64\times64\times64$ &  3D heat propagation equation. \\

        \texttt{axpy} & $32768$ &  Generic vector-vector addition. \\
        
        \texttt{merge sort} & $65536$ &  Merge sort algorithm. \\
    
        \bottomrule
    
    \end{tabular}
    \label{tab:kern_desc}
    \end{center}
    \vspace{-2em}
\end{table}


\section{Results}

\begin{table*}[ht!]
\caption{Total runtime in cycles for each kernel at variable memory latency.}
\begin{center}
\resizebox{\linewidth}{!}{
\begin{threeparttable}


\begin{tabular}{|l|ccc|ccc|ccc|ccc|}

\hline

Kernel & \multicolumn{3}{c|}{$gemm_{128}$} & \multicolumn{3}{c|}{$gesummv_{512}$} & \multicolumn{3}{c|}{$heat3d_{64}$} & \multicolumn{3}{c|}{$mergesort_{65536}$} \\ 

\hline

DRAM Latency & 200 & 600 & \textbf{1000} & 200 & 600 & \textbf{1000} & 200 & 600 & \textbf{1000} & 200 & 600 & \textbf{1000} \\ 

\hline

Baseline & 2.03e6 & 2.24e6 & 2.45e6 & 4.93e5 & 6.38e5 & 9.16e5 & 2.00e6 & 4.60e6 & 7.21e6 & 6.94e6 & 7.98e6 & 9.05e6 \\
\% DMA   & \multicolumn{1}{l}{7.3\%}  & \multicolumn{1}{l}{16.0\%}   & \multicolumn{1}{l|}{\textbf{23.2\%}} 
         & \multicolumn{1}{l}{1.4\%}  & \multicolumn{1}{l}{23.5\%} & \multicolumn{1}{l|}{\textbf{46.3\%}} 
         & \multicolumn{1}{l}{36.3\%} & \multicolumn{1}{l}{71.9\%} & \multicolumn{1}{l|}{\textbf{80.8\%}} 
         & \multicolumn{1}{l}{17.7\%} & \multicolumn{1}{l}{29.2\%} & \multicolumn{1}{l|}{\textbf{38.3\%}} \\ 

\hline

IOMMU & 2.12e6 & 2.50e6 & 2.89e6 & 5.20e5 & 1.08e6 & 1.70e6 & 2.84e6 & 7.09e6 & 1.13e7 & 7.67e6 & 1.08e7 & 1.44e7 \\
      & \multicolumn{1}{l}{11.1\%} & \multicolumn{1}{l}{24.6\%} & \multicolumn{1}{l|}{\textbf{34.5\%}} 
      & \multicolumn{1}{l}{6\%}    & \multicolumn{1}{l}{54\%}   & \multicolumn{1}{l|}{\textbf{70.4\%}} 
      & \multicolumn{1}{l}{54.9\%} & \multicolumn{1}{l}{78.9\%} & \multicolumn{1}{l|}{\textbf{84.8\%}} 
      & \multicolumn{1}{l}{27\%}   & \multicolumn{1}{l}{63.4\%} & \multicolumn{1}{l|}{\textbf{82.6\%}} \\ 
      
\hline

IOMMU+LLC & 2.04e6 & 2.25e6 & 2.47e6 & 4.95e5 & 6.45e5 & 9.29e5 & 2.05e6 & 4.68e6 & 7.30e6 & 6.96e3 & 8.00e6 & 9.07e6 \\
    & \multicolumn{1}{l}{7.7\%}  & \multicolumn{1}{l}{16.4\%} & \multicolumn{1}{l|}{\textbf{23.7\%}} 
    & \multicolumn{1}{l}{1.5\%}  & \multicolumn{1}{l}{24.1\%} & \multicolumn{1}{l|}{\textbf{46.9\%}} 
    & \multicolumn{1}{l}{37.8\%} & \multicolumn{1}{l}{72.2\%} & \multicolumn{1}{l|}{\textbf{81.0\%}} 
    & \multicolumn{1}{l}{22.4\%} & \multicolumn{1}{l}{29.5\%} & \multicolumn{1}{l|}{\textbf{38.6\%}} \\ 

\hline

\end{tabular}


\begin{tablenotes}
\footnotesize

\end{tablenotes}

\end{threeparttable}
}

\label{tab:results}
\end{center}
\vspace{-2em}
\end{table*}

Using the platform and benchmarks described above, we present the performance impact of shared virtual addressing in this section. First, we present the benefits of shared addressing thanks to zero-copy offloading at the application scale. Then, we focus on its impact on the accelerator runtime at the compute kernel scale.

\subsection{Zero-copy offloading}

To show the benefits of shared virtual addressing on heterogeneous applications, we compare the execution time of a single precision \texttt{axpy} operation. We consider three cases:  CVA6 executes the kernel; CVA6 copies the data to shared \gls{dram} and the Snitch cluster executes the kernel; CVA6 creates \gls{iova} mapping and Snitch cluster executes the kernel in a zero-copy fashion. We select a relatively small problem size of $32.768$ elements per vector (corresponding to 16~input~pages) in order to present the different overheads accurately and prevent the compute time from dominating the whole execution. We split the accelerated application into three regions: the time required to copy or map data to the cluster, the overhead caused by triggering execution and synchronizing using the OpenMP target fork/join model, and the computation on the device.

\begin{figure}[t]
\centering
\includegraphics[width=1\columnwidth]{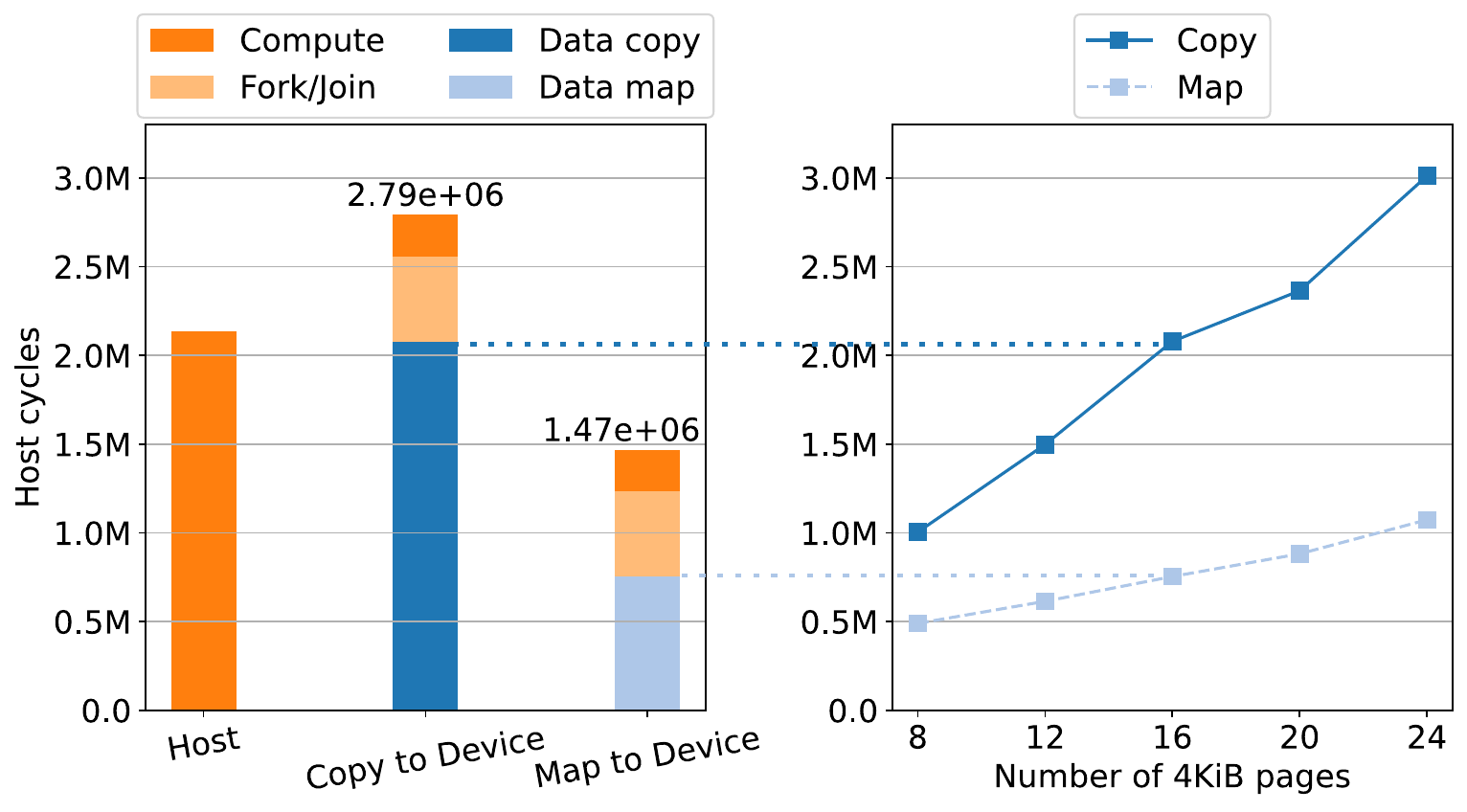}
\caption{(Left) \texttt{axpy}$_{32.768}$ breakdown for three scenarios. Host-only execution; Data copy and device execution;  Data mapping and device execution. (Right) Time spent copying or mapping data of different input sizes.}
\label{fig:axpy-time}
\end{figure}

In the left half of Figure~\ref{fig:axpy-time}, we show the runtime breakdown for all three cases. First, the computation part is always faster on the cluster since it contains eight \glspl{pe} and does not need to handle kernel-related hardware interrupts. Then, we note that offloading can be more expensive with explicit data copy than a regular host execution. This is particularly true for an \texttt{axpy} kernel that is heavily memory-bound. However, we also note that although page mapping requires at most 24~bytes (three-page table entries) per every \SI{4}{\kibi\byte} of input, a consequent overhead of a few hundred thousand cycles remains in the third case. This is partly due to the communication with the Linux kernel module via \texttt{ioctl}. Overall, we measure in Figure~\ref{fig:axpy-time} that zero-copy offloading is $47\%$ faster than copy-based offloading. On the right side of Figure~\ref{fig:axpy-time}, we present the explicit copy and zero-copy runtime with increasing problem sizes. As expected, zero-copy offloading also scales better with problem size than an explicit copy.

\begin{figure}[t]
\centering
\includegraphics[width=0.96\columnwidth]{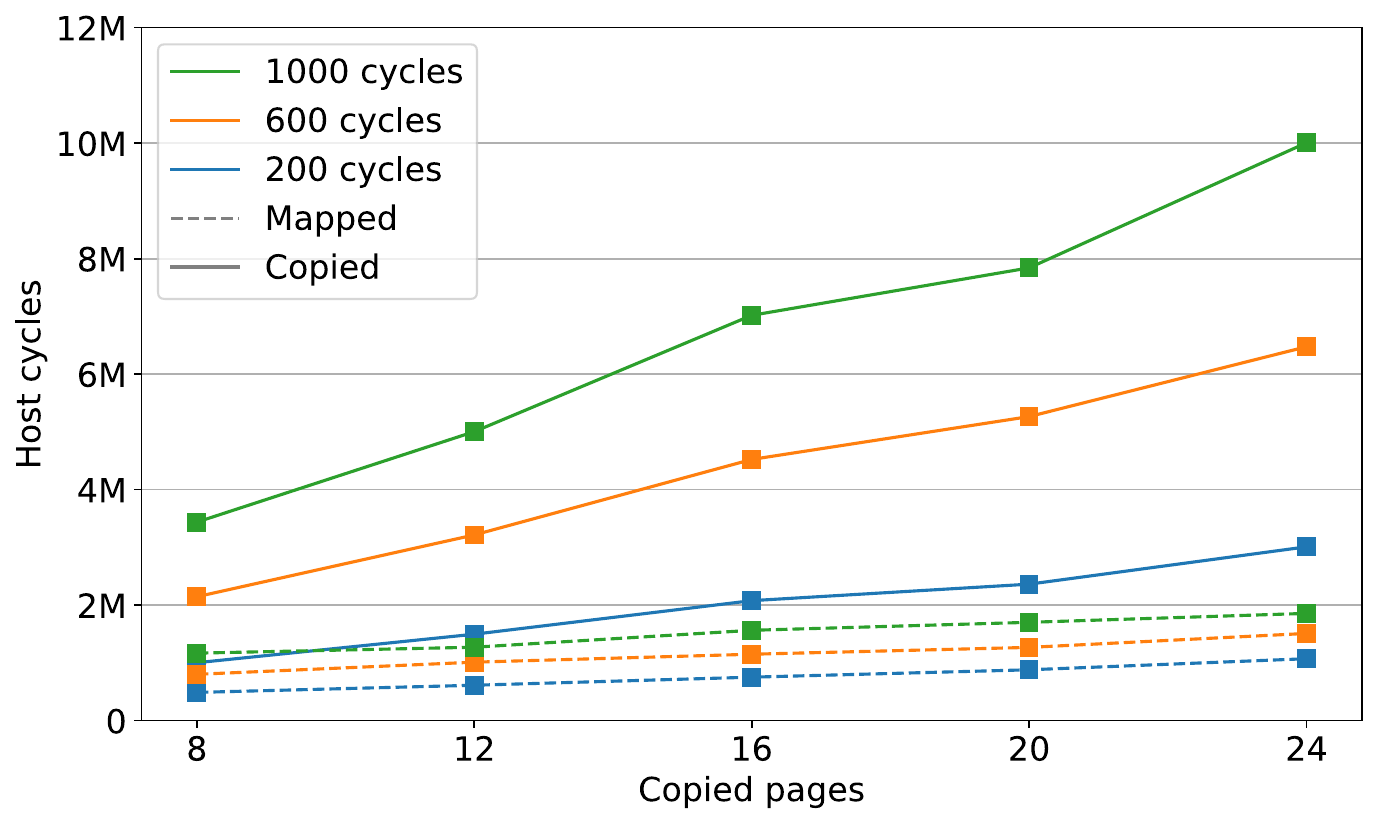}
\caption{Data copying and mapping time with input size and different \gls{dram} latencies.}
\label{fig:map-time}
\end{figure}

However, data copy overhead does not only depend on the input size. In Figure~\ref{fig:map-time}, we compare copying and mapping for increasing \gls{dram} latencies. We note that copying 16~pages (\SI{64}{\kibi\byte}) of data to the physically contiguous portion of the \gls{dram} is $3.4\times$ slower when the memory latency raises from 200~cycles to 1000~cycles. However, for the same variation in latency, the time required to map the IO page entries time is multiplied by $2.1\times$. This is due to IO-mapping accessing data structures likely to be present in the CVA6 data cache.

\subsection{Device execution and IOMMU overhead}

The previous section shows that shared virtual addressing drastically reduces offloading overhead in heterogeneous applications. Nevertheless, using IO virtual addresses requires the accelerator to translate each access to \gls{dram}. This IO page table walking is a sequential operation that may significantly increase the time spent on the accelerator, especially under high memory latency.

\begin{figure}[]
\centering
\includegraphics[width=1\columnwidth]{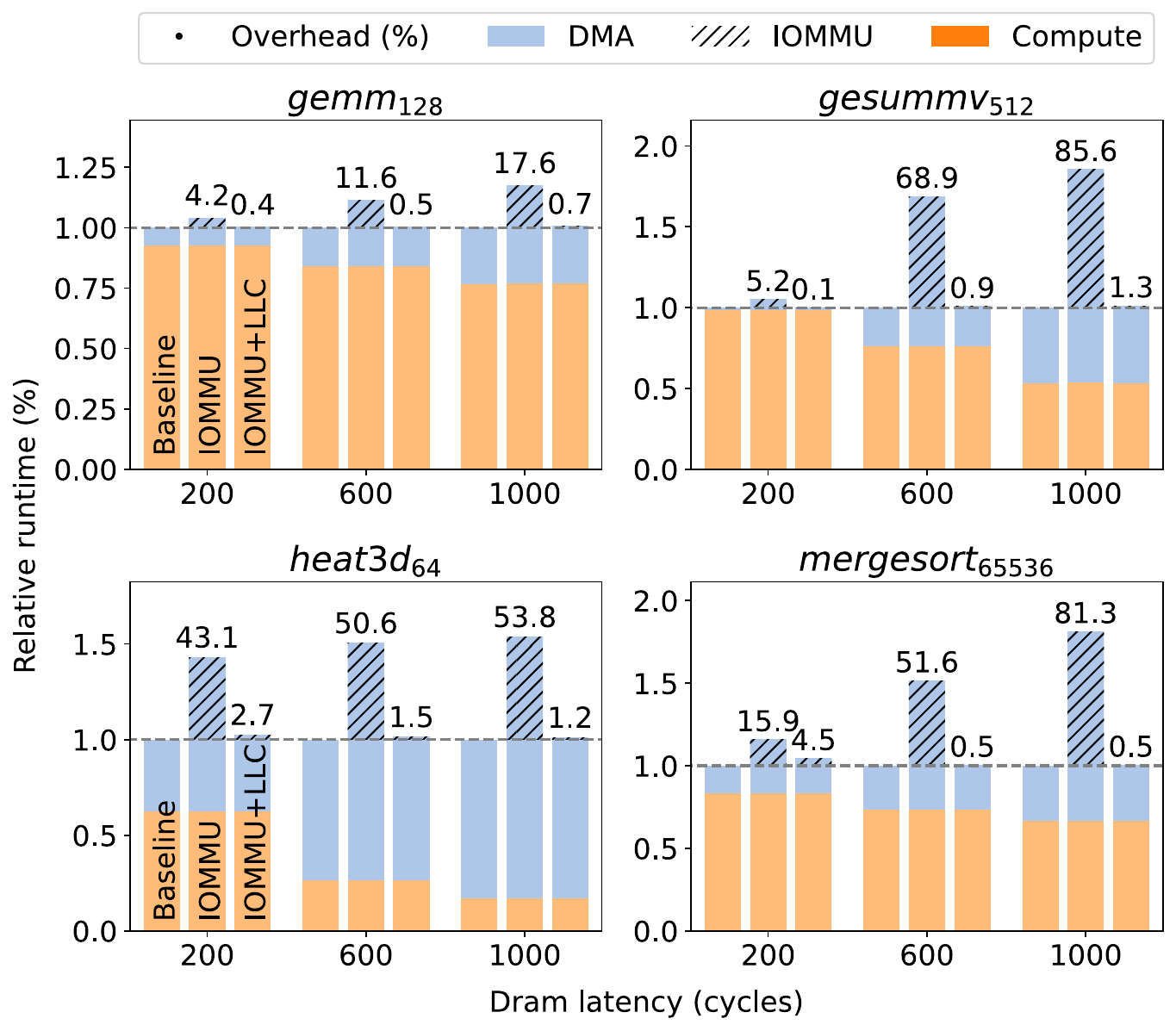}
\caption{Kernel execution for different \gls{dram} latencies with three configurations. With IOMMU disabled, with IOMMU enabled and LLC disabled, and with IOMMU enabled and LLC enabled.}
\label{fig:kernels-2}
\end{figure}

In this section, we measure the runtime on the accelerator for the four selected kernels under different \gls{dram} latencies. We do not measure offloading or synchronization time with the host, and we propose a breakdown into two regions. The \gls{dma} region measures the cycles where the cores are busy waiting for data transfers. The compute region measures the rest of the cycles to complete the kernel on the accelerator. Note that, thanks to double-buffering and a dedicated \gls{dma}-engine, when the kernel is compute bound, the \gls{dma} region tends to zero even if megabytes of data are transferred. We take the total runtime without IOMMU as the baseline, and we plot the relative runtime for all configurations in Figure~\ref{fig:kernels-2}. For IOMMU-enabled configurations, we add the percentage overhead on the plot, and we display for all experiments the absolute measurements in Table~\ref{tab:results}. We first note that for all kernels, the time spent waiting for \gls{dma} increases with memory latency, even without IOMMU. Naturally, this increase is smaller for kernels with high arithmetic intensity such as \texttt{gemm} (up to $23,2\%$) than for more memory intensive kernels such as \texttt{heat3d} (up to $80.8\%$). When using IO virtual addresses, the latency of a single \gls{dma} transfer may increase by $300\%$ since an IOTLB miss requires three new additional memory accesses. In case a \gls{dma} transfer is larger than a page, it will be split into multiple bursts (according to the \gls{axi} specifications), and every burst causing IOTLB misses may reduce the effective memory bandwidth for the \gls{dma}-engine. In Figure~\ref{fig:kernels-2}, we show that this reduced effective bandwidth creates an overhead of $81.3\%$ compared to baseline for the \texttt{heat3d} kernel with 1000~cycles of \gls{dram} latency. On the other hand, thanks to high data reuse, this translation overhead is of $17.6\%$ in the \texttt{gemm} kernel under the same conditions.

\begin{lstlisting}[belowskip=-15pt, float, caption={
Self invalidation based coherency.},label={lst:invalidate}, language=C, style=CStyle]
a = malloc(n_bytes)
prepare_input(a)
flush_l1()
flush_last_level_cache()
a_iova = create_iommu_mapping(a, n_bytes)
flush_l1()
#pragma omp target device(1) map(to: a_iova)
  device_kernel(a_iova + LLC_BYPASS_OFFSET)
\end{lstlisting}

To face the significant overhead caused by IO virtual address resolution, we support the idea that a shared last-level cache is necessary for heterogeneous architectures using shared virtual addressing. Usually, scratchpad-based accelerators rely on \gls{dma} engines to amortize long latency accesses to \gls{dram}. The presence of an \gls{llc} may lower the accelerator's bandwidth by reducing long bursts to the length of a cache line. However, due to the presence of the \gls{iommu}, the accelerator now also needs to access page table entries at a low granularity with low latency. To facilitate such accesses, we propose to leverage a shared \gls{llc}, caching only host and \gls{iommu} \gls{ptw} memory transactions. We show in Figure~\ref{fig:kernels-2} and Table~\ref{tab:results} the effect of this \gls{llc} on the performance. For all selected kernels and at all memory latency, the \gls{iommu} overhead is now lower than $2\%$ of the total runtime.

To ensure that the \gls{llc} is only used by the host subsystem, the \gls{iommu}, and not the \gls{dma} engine, we use a pair of demux and mux visible in Figure~\ref{fig:archi:soc}. These architectural blocks allow to remap the same \gls{dram} addresses to two bus addresses separated by a fixed offset. With this offset, the \gls{dma} engine can access input data in bursts larger than the cache line and write output data without invalidating unrelated host data. Similar bypass regions can be implemented on \gls{noc} platforms~\cite{heterogeneous_esp_2022}. However, using uncached \gls{dma} transactions requires the host to self-invalidate data from the \gls{llc} when the device needs to access it as shown in Listing~\ref{lst:invalidate}. Note that by flushing the LLC before mapping the IO virtual addresses, we maximize the \gls{llc} hits for the device IOMMU.

\subsection{Page table walking time and \gls{llc} impact}

\begin{figure}[]
\centering
\includegraphics[width=0.8\columnwidth]{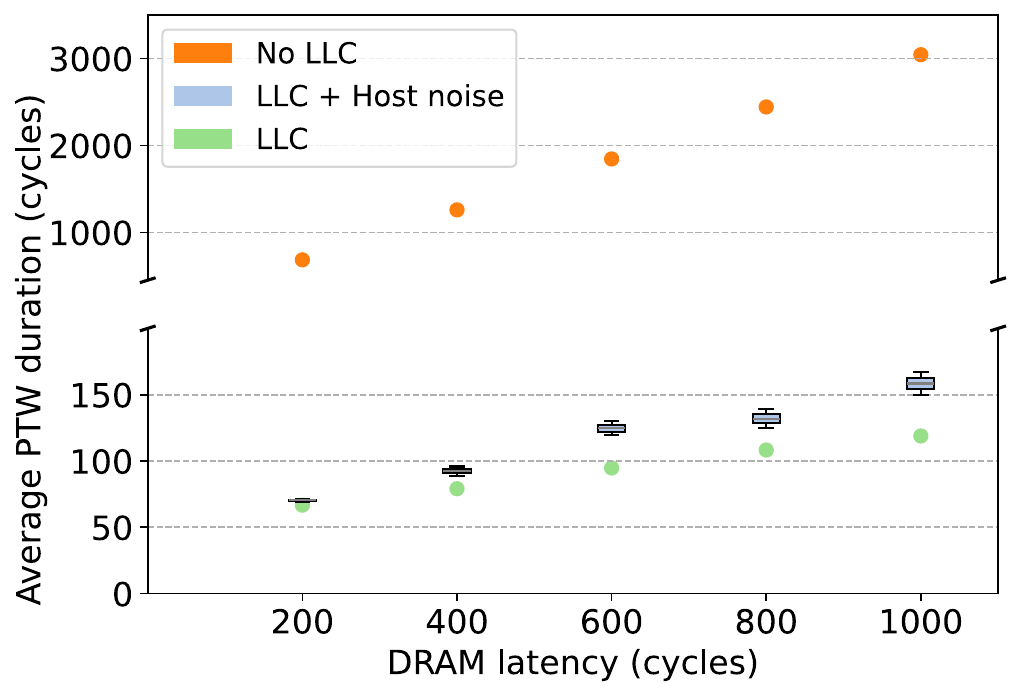}
\caption{Average IOMMU page table walk time with and without \gls{llc} and host interference for increasing \gls{dram} latencies.}
\label{fig:plot_box}
\end{figure}

In this part, we confirm the benefits of sharing a last-level cache for the \gls{iommu} and host. Similarly to the previous section, we focus on runtimes and overhead on the device. We measure the average IO \gls{ptw} time when running the \texttt{axpy} kernel with the shared \gls{llc} when it is concurrently stressed by host memory traffic. We create this interference by issuing a synthetic random memory from the host during the accelerator's execution. 
Figure~\ref{fig:plot_box} shows the average IO page table walk time under varying \gls{dram} latencies. We observe that using a shared \gls{llc} between the accelerator and host domains significantly improves the IOMMU performance by reducing the \gls{ptw} time by $15\times$ on average. Indeed, thanks to the \gls{llc}, the average page table walk time does not exceed $200$ clock cycles, even for 1000~cycles of \gls{dram} latency. The page table entries have a high chance of being cached in the \gls{llc} as the mapping operation is performed by the host core right before offload, following the offload model shown in Listing~\ref{lst:invalidate}. This reduces the probability of compute and energy expensive \gls{llc} misses.

However, Figure~\ref{fig:plot_box} shows that the activity of the host may also affect the IOMMU address translation performance due to interference on the system bus and concurrent evictions in the \gls{llc}. We measure in average a \SI{20}{\percent} slowdown of the \gls{ptw} time when CVA6 frequently accesses memory. Although such interference must be taken into consideration when evaluating the feasibility of implementing IOMMU-based shared virtual memory heterogeneous systems, the \gls{llc} still provide considerable performance improvements without requiring any custom support in the \gls{iommu}.


\section{Related works}

Heterogeneous \glspl{soc}, often referred to as \glspl{apu}, early adopted shared virtual addressing via \gls{iommu} hardware units to enhance programmability and memory usage. Several works evaluated the key performance bottlenecks~\cite{amd_benchmarking_2016}~\cite{host_congestion_2022} induced by the additional IO page walking required to translate IO virtual addresses to physical addresses. For example, \glspl{gpu} typically rely on deep cache-hierarchy and latency tolerant \gls{simt} to maximize compute utilization. Upon L2-cache misses, \glspl{gpu} access \gls{dram} via a shared \gls{llc} with the \gls{cpu}. Conversely, \gls{mimd} accelerator clusters typically rely on scratchpad memories and \gls{dma}-engines to refill contiguous data chunks from \gls{dram}~\cite{hero_archi_2018}. In this scenario, different \gls{iommu} architectures have been proposed over the years, such as multi-level IO\glspl{tlb}~\cite{multi_lvl_tlb_2017}, enabling sharing of commonly accessed \glspl{pte} among different accelerators, significantly increasing performance on shared data access patterns.

On the other end of the spectrum, previous works also explored IOMMU-less solutions to enable shared virtual addressing to improve flexibility and reduce area and power consumption. A former study from Kurth et al.~\cite{hero_scalable_iommu_2018} proposes a hybrid architecture for heterogeneous systems-on-chip based on programmable accelerators, in which page table walking and prefetching are executed by software threads running on the accelerator cluster itself. This solution enables high versatility by tuning the proportion of page walker threads per compute thread for each kernel. Nevertheless, on systems as the one described in section~\ref{sec:platform}, software-based page table walking could waste precious computational resources such as the double-precision \gls{fpu} coupled to each \gls{pe}. To reduce energy consumption,~\cite{active_forwarding_2018} proposes a hardware-managed active forwarding from the host data cache to the accelerator's \gls{spm}. This solution is reported to reduce energy utilization significantly for simple dataflow accelerators. However, active forwarding is limited to problem sizes that can entirely fit within the (typically limited) on-chip accelerator scratchpad memory. While these limitations in terms of flexibility can be tolerated on certain \glspl{dsa} with highly predictable execution patterns, accelerators based on programmable \glspl{pe} justify the use of dedicated hardware \gls{iommu} between the \gls{pmca} and the system bus.

Within the context of \gls{iommu}-based approaches, Ben-Yehuda et al.~\cite{iommu_prize_2007} identified two types of performance overheads: \gls{cpu} utilization and memory bandwidth. In the case of low memory bandwidths, hardware solutions have been proposed to limit the IO\gls{tlb} wall. \gls{tlb} coalescing~\cite{colt_coalescing_2012} can be used in the host \gls{mmu} to reduce the number of \glspl{ptw} without the need of superpages. This method can be similarly applied to \glspl{iommu}. In~\cite{iotlb_coalescing_2018}, the authors propose to exploit the locality of IO page table entries in cache lines for efficient coalescing. To adress the problem of \gls{cpu} overhead. In this case, software solutions can be used to reduce the cost of mapping and un-mapping \gls{dma} buffers. In~\cite{iommu_allocator_2018}, the authors propose a \gls{dma} allocator specially designed to reuse \gls{iommu} mappings.

In this work, we focus on the RISC\nobreakdash-V \gls{iommu} proposed in~\cite{pinto_iommu_2023} coupled with a scratchpad-based programmable accelerator cluster with \gls{dma} engine. We benchmark the page translation overhead on the accelerator under different memory latencies to exhibit the high cost of shared virtual addressing. We also show that by integrating a last-level cache shared by the \gls{iommu} and the host, page walking overhead is highly reduced, making shared virtual addressing suitable without custom IO\gls{tlb} coalescing or prefetching. However, to avoid lowering the accelerator's bandwidth, the shared \gls{llc} must bypass \gls{dma} transactions. This bypass is implemented in platforms such as ESP~\cite{heterogeneous_esp_2022} to enable high bandwidth within network-on-chips. Our work shows that such memory hierarchy is also required to implement shared virtual addressing on heterogeneous platforms efficiently.


\section{Conclusion}

In this work, we evaluate the performance overhead of shared virtual addressing on an open source RISC\nobreakdash-V heterogeneous \gls{soc}. Our prototype platform contains a 64\nobreakdash-bit CVA6 core running Linux and an 8\nobreakdash-cores Snitch cluster. We implement several Linux userspace heterogeneous kernels using OpenMP and benchmark them under variable off-chip memory latencies. We show that \gls{iommu}-based \gls{iova} translation can significantly reduce the effective memory bandwidth for the accelerator and contribute to up to $17.6\%$ of the accelerator's execution time for a \texttt{gemm} kernel. However, we also show that this issue can be mitigated without any changes to the \gls{iommu} by integrating a last-level cache shared by the host and the IO page table walker. While classical programmable manycore accelerator architectures rely on direct \gls{dram} access via \gls{dma} and do not need an \gls{llc}, we show that this hardware unit reduces address translation overhead to less than $2\%$ for all kernels benchmarked, even under high memory latency. We also propose to bypass the shared cache for the device \gls{dma} as it may lower the effective \gls{dma} bandwidth under large bursts. In these conditions, and using the open source RISC\nobreakdash-V \gls{iommu} from~\cite{pinto_iommu_2023}, we show that shared virtual addressing and zero-copy offloading is suitable for heterogeneous RISC\nobreakdash-V systems-on-chip.


\section{Acknowledgment}

This work has been supported in part by the EPI SGA2 project, which received funding from the European High-Performance Computing Joint Undertaking (JU) (Specific Grant Agreement No 101036168), and the TRISTAN project, which received funding from the HORIZON KDT-JU program (Grant Agreement No 101095947).

\bibliographystyle{ieeetr}
\bibliography{refs} 

\end{document}